\documentclass[notitlepage,aps,11pt]{revtex4-1}
\expandafter\let\csname equation*\endcsname\relax
\expandafter\let\csname endequation*\endcsname\relax
\expandafter\let\csname eqnarray*\endcsname\relax
\expandafter\let\csname endeqnarray*\endcsname\relax
\usepackage{amsmath}
\usepackage{amssymb}
\usepackage{graphicx}
\newcommand{\frot}{\frac{1}{2}}

\begin{document}
\title{Gravitational properties of light - The emission of counter-propagating laser pulses from an atom}
\author{Dennis R\"atzel, Martin Wilkens, Ralf Menzel}
\address{University of Potsdam, Institute for Physics and Astronomy\\
Karl-Liebknecht-Str. 24/25, 14476 Potsdam, Germany}
%\ead{raetzel@uni-potsdam.de}
\begin{abstract}
The gravitational field of a laser pulse, although not detectable at the moment, comes with a peculiar feature which continues to attract attention; cause and effect propagate with the same speed, that of light. A particular result of this feature is that the gravitational field of an emitted laser pulse and the gravitational effect of the emitter's energy-momentum change are intimately entangled. In this article, a specific example of an emission process is considered - an atom, modeled as a point mass, emits two counter-propagating pulses. The corresponding curvature and the effect on massive and massless test particles is discussed. A comparison is made with the metric corresponding to a spherically symmetric massive object that isotropically emits radiation - the Vaidya metric.


\noindent PACS numbers: 04.20.-q, 42.55.-f, 42.60.Jf, 42.62.-b, 42.55.Ah
%Relativity - general relativity - classical: 04.20.-q
%Lasers, 42.55.-f
%Laser radiation - characteristics, 42.60.Jf
%Lasers, applications of, 42.62.-b
%Lasers, general theory of, 42.55.Ah
%ams: Quantum theory - gravitational interaction 81V17
%ams: Quantum optics 81V80
%ams: Optics, electromagnetic theory - waves and radiation 78A40
%ams: Optics, electromagnetic theory - lasers etc 78A60
%ams: Relativity and gravitational theory - Approximation procedures, weak fields 83C25
%ams: Relativity and gravitational theory - Gravitational waves 83C35
%ams: Relativity and gravitational theory - Electromagnetic fields 83C50 
\end{abstract}
%\ams{83C25, 83C50, 83C35, 78A60, 78A40, 81V80}

%\keywords{linearized gravity, laser pulses, electromagnetic field, gravitational field}

\maketitle

%\newpage
\section{Introduction}
\label{sec:introduction}

%The gravitational field of a laser pulse, although not detectable at the moment, comes with a peculiar feature which continues to attract attention; cause and effect propagate with the same speed, that of light.

In a seminal paper \cite{Tolman1931} by Tolman et al., a single light pulse of finite lifetime was modeled and the corresponding gravitational field was derived as a perturbation of Minkowski spacetime in the framework of linearized gravity. In \cite{Bonnor2009} and \cite{Raetzel2016pulse}, it was argued that the gravitational field of the pulse is due only to its emission and absorption. 
In \cite{Raetzel2016pulse}, it was shown that the exact process in which a laser pulse is emitted can be disregarded, as long as one is only interested in the gravitational field close to the pulse trajectory. However, in general, the change in the emitter's energy-momentum accompanying the emission of the pulse induces a gravitational effect that is intimately entangled with the gravitational effect of the creation of the pulse. In particular, like the laser pulse and the gravitational effect of its emission, the gravitational effect of the emitter's energy-momentum change propagates with the speed of light.

In this article, a specific example of an emission process will be considered - an atom, modeled as a point mass, emits two counter-propagating pulses. This model respects the continuity equation of general relativity $\partial_\mu T^{\mu\nu}=0$ and the results are valid for all spacetime points. The model will be introduced and the corresponding gravitational field will be derived in Section \ref{sec:pulse}. In Section \ref{sec:curvature}, the curvature and the tidal forces will be given. In Section \ref{sec:acc} and \ref{sec:masslesstest}, the acceleration of test particles in the bi-pulse metric will be discussed. 

The situation of two laser pulses emitted from a point particle shares certain features with the spacetime corresponding to a spherically symmetrical massive object that emits radiation isotropically. The corresponding solution of the full Einstein equations is called the Vaidya metric. In Section \ref{sec:compvaidya}, the relation to the Vaidya metric will be clarified and the acceleration of a massive test particle in this spacetime will be discussed.

\section{Two laser pulses emitted by a point mass}
\label{sec:pulse}

\begin{figure}[h]
%\hspace{2.2cm}
\includegraphics[width=12cm,angle=0]{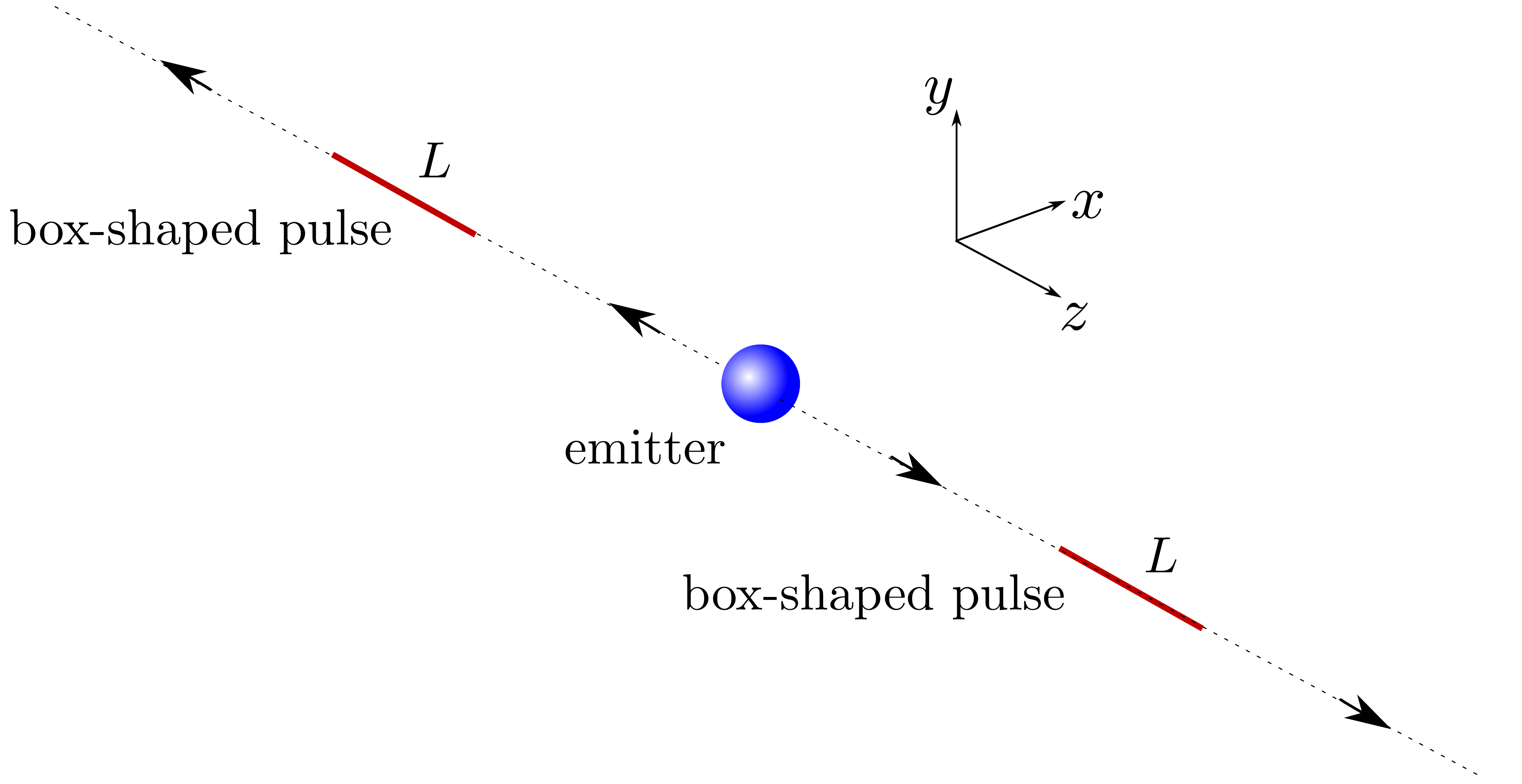}
\caption{The laser pulses are modeled as pulses of electromagnetic radiation of length $L$, traveling from the emitter along the $z$-axis in the negative and positive $z$-direction. The extension of the pulses in the transverse $x/y$-directions is assumed to be negligible in comparison to their length.
 \label{fig:pulse}}
\end{figure}

In this section, we will present the model for the emission of two counter-propagating laser pulses from a point particle. The situation is illustrated in Figure \ref{fig:pulse}. We start with the separation of the metric into the background Minkowski metric $\eta_{\mu\nu}=\mathrm{diag}(-,+,+,+)$ and a metric perturbation $h_{\mu\nu}$
\begin{equation}\label{eq:metricsplit}
	g_{\mu\nu} = \eta_{\mu\nu} + h_{\mu\nu}\,,
\end{equation}
where $|h_{\mu\nu}|\ll 1$ is assumed to hold. We assume the Lorentz gauge condition
\begin{equation}\label{eq:lorentzgauge}
	\partial^\mu \left(h_{\mu\nu} - \frac{1}{2}\eta_{\mu\nu} {h_\alpha}^\alpha\right)=0\,,
\end{equation}
and the Einstein field equations become, in first order in the metric perturbation (see \cite{Misner1973} page 438)
\begin{equation}\label{eq:linearizedeinstein}
	\left[\frac{1}{c^2}\frac{\partial^2}{\partial t^2} -\frac{\partial^2}{\partial x^2}
	- \frac{\partial^2}{\partial y^2}
	- \frac{\partial^2}{\partial z^2} 
	\right] h_{\mu\nu} = \frac{16\pi G}{c^4}
	\left(T_{\mu\nu}-\frac{1}{2}\eta_{\mu\nu} {T_\alpha}^\alpha\right)\,,
\end{equation}
where $T_{\mu\nu}$ is the energy-momentum tensor of the two pulses and the point mass. 
We assume the point mass to be at rest in the lab frame and the emission to happen at $z=0$, to start at $t=0$ and to end at $t=L/c$, where $L$ the length of the laser pulses. The only non-zero component of the energy momentum tensor for the emitting point particle is given as
\begin{equation}\label{eq:energymomentum}
	\nonumber  T^s_{00}=\left\{
	\begin{array}{ccc}
		mc^2\delta^{(3)}(\vec r) &:& ct\le 0\\
		\left(mc^2-2\epsilon\frac{ct}{L}\right)\delta^{(3)}(\vec r)   &:& 0 < ct < L\\ 
		\left(mc^2-2\epsilon\right)\delta^{(3)}(\vec r)  &:& L \le ct\,.
	\end{array}
	\right.		
\end{equation}
where $m$ is the mass of the point particle before the emission and $\epsilon$ is the energy, $\hbar \omega$, of each laser pulse.

We model each pulse as a pulse of electromagnetic radiation, traveling from the emitter along the $z$-axis, with finite extension (pulse length) $L$ in the direction of propagation, but negligible extension $\Delta(z)$ in the transverse $x/y$-directions, $\Delta(z)\ll L$ (see figure \ref{fig:pulse}) such that we can well approximate the energy density as proportional to $\delta(x)\delta(y)$. All measures refer to a laboratory frame where the emitting system is at rest before emission of the pulse. The first pulse moves in the positive $z$-direction and the second pulse moves in the negative $z$-direction. Hence, in the coordinates $(ct,x,y,z)$, the terms in the energy momentum tensor due to the pulses have the non-zero components $T^{+}_{00}=T^{+}_{zz}=-T^{+}_{z0}=-T^{+}_{0z}=T^+$ and $T^{-}_{00}=T^{-}_{zz}=T^{-}_{z0}=T^{-}_{0z}=T^-$. We consider only circularly polarized laser pulses, which means that the energy momentum tensor is constant over the full extension of the pulses along the $z$-axis. Then, the functions $T^{+}$ and $T^{-}$ have the following form:
\begin{equation}\label{eq:Tmunufunctions}
	 \begin{array}{llll}
	 	\textrm{For }ct\le 0: & 
	 	T^+=0\,, &T^-=0\\
	 	\textrm{For } ct > 0  \,\, \mathrm{and} \,\,  z>0: & 
	 	T^+= \frac{\epsilon}{L} \delta(x)\delta(y)\chi(ct-z)\, &
	 	T^-= 0\\
	 	\textrm{For }ct > 0  \,\,  \mathrm{and} \,\,  z<0: & 
	 	T^+= 0\,&
	 	T^-= \frac{\epsilon}{L} \delta(x)\delta(y)\chi(ct+z)
	 \end{array}
\end{equation}
where $\chi$ is a characteristic function (normalized $\chi^2=\chi$) which -- for any given time $t$ -- encodes the momentary extension and location of the laser pulses on the $z$-axis. It is explicitly given as 
\begin{equation}
	\nonumber  \chi(x)=\left\{
	\begin{array}{ccc}
		0 &:& x\le 0\\
		1  &:& 0 < x < L\\ 
		0  &:& L \le x\,.
	\end{array}
	\right.		
\end{equation}

The linearized Einstein equation (\ref{eq:linearizedeinstein}) can be solved by the retarded potential as (see \cite{Misner1973} page 445)
\begin{equation}\label{eq:retarded}
	h_{\mu\nu}(x,y,z,t)
	=
	\frac{4 G}{c^4} \int  \frac{
	\left(T_{\mu\nu}-\frac{1}{2}\eta_{\mu\nu}{T_\alpha}^\alpha\right)(x',y',z',t_\mathrm{ret})}{\sqrt{(x-x')^2+(y-y')^2+(z-z')^2}} dx'dy'dz'\,,
\end{equation}
with $t_\mathrm{ret}$ the retarded time, $t_\mathrm{ret}=t-\sqrt{(x-x')^2+(y-y')^2+(z-z')^2}/c$. The retarded potential for the energy momentum tensor of the point particle is easily calculated. The pulses define world sheets restricted to the $t$-$z$-plane starting from the emission points. 
\begin{figure}[h]
%\hspace{2cm}
\includegraphics[width=10cm,angle=0]{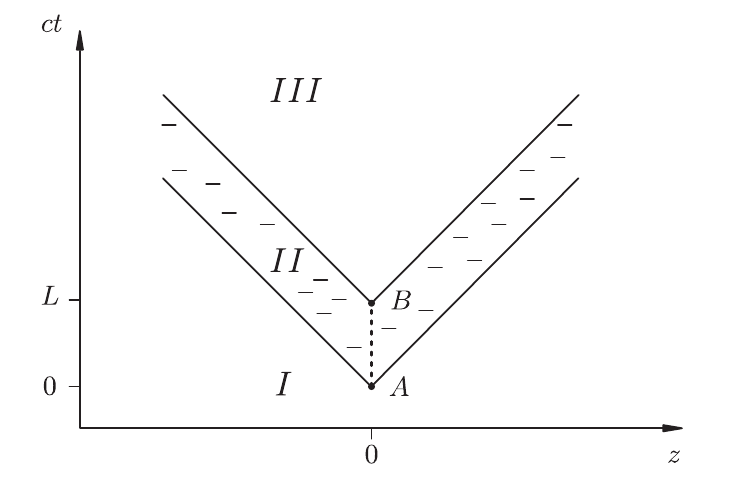}
\caption{ The pulses define world sheets restricted to the $t$-$z$-plane. The world sheets touch along the world line of the emitter between the points A and B which correspond, respectively, to the start of the pulse emission and the end of the emission. The future directed light cones of A and B define the spacetime regions $I$-$III$ with qualitatively different metric perturbations. Region $I$ is the pre-emission cone, region $II$ is the emission cone and region $III$ is the post-emission cone.
 \label{fig:Regions}}
\end{figure}
The retarded solution (\ref{eq:retarded}) for the pulses is given by the intersection of the backward light cone of the spacetime point $x$ with the world sheets. Hence, there are three different regions defined by the future directed light cones of the emission process. A spacetime diagram illustrating this situation is given in Figure \ref{fig:Regions}. We call these regions the pre-emission cone, the emission cone and the post-emission cone, respectively. 

From the retarded potential (\ref{eq:retarded}), we obtain the metric perturbation
\begin{equation}\label{eq:gmunudecay}
		h_{\mu\nu}=h^{s}_{\mu\nu}+h^{+}_{\mu\nu}+h^{-}_{\mu\nu}
\end{equation}
where the three terms $h^{s}_{\mu\nu}$, $h^{+}_{\mu\nu}$ and $h^{-}_{\mu\nu}$ correspond to the point particle, the laser pulse propagating in positive $z$-direction and the laser pulse propagating in negative $z$-direction, respectively. In the coordinates $(ct,x,y,z)$, they have the non-zero components $h^{s}_{00}=h^{s}_{xx}=h^{s}_{yy}=h^{s}_{zz}=h^s$, $h^{+}_{00}=h^{+}_{zz}=-h^{+}_{z0}=-h^{+}_{0z}=h^+$ and $h^{-}_{00}=h^{-}_{zz}=h^{-}_{z0}=h^{-}_{0z}=h^-$.
The function $h^{s}$, $h^{+}$ and $h^{-}$ have the following form:
%\begin{eqnarray}\label{eq:gfunctions}
%	 \nonumber h^{s}&=&\left\{
%	\begin{array}{ccc}
%		 \frac{mc^2}{2c^4}\frac{4G}{r} &:& ct-r\le 0\\
%		 \frac{1}{2c^4} \left(mc^2-2\epsilon\frac{ct-r}{L}\right)\frac{4G}{r}  &:& 0 < ct-r < L\\ 
%		 \frac{1}{2c^4} \left(mc^2-2\epsilon\right)\frac{4G}{r} &:& L \le ct-r
%	\end{array}\right.
%	 \\
%	 \nonumber h^{+}&=&\left\{
%	\begin{array}{ccc}
%		 0 &:& ct-r\le 0\\
%		  \frac{\epsilon}{c^4}\frac{4G}{L} \ln\frac{ct-z}{r-z} &:& 0 < ct-r < L\\ 
%		 \frac{\epsilon}{c^4}\frac{4G}{L} \ln\frac{ct-z}{ct-L-z} &:& L \le ct-r
%	\end{array}\right.	 \\
%	 \nonumber h^{-}&=&\left\{
%	\begin{array}{ccc}
%		 0 &:& ct-r\le 0\\
%		  \frac{\epsilon}{c^4}\frac{4G}{L} \ln\frac{ct+z}{r-z} &:& 0 < ct-r < L\\ 
%		 \frac{\epsilon}{c^4}\frac{4G}{L} \ln\frac{ct+z}{ct-L+z} &:& L \le ct-r
%	\end{array}\right.
%\end{eqnarray}
\begin{equation}\label{eq:gfunctions}
	 \begin{array}{lllll}
	 	\textrm{For }ct-r\le 0: & h^s=\frac{mc^2}{2c^4}\frac{4G}{r}\,, &
	 	h^+=0\,, &h^-=0\\
	 	\textrm{For }0 < ct-r < L: & h^s=\frac{1}{2c^4} \left(mc^2-2\epsilon\frac{ct-r}{L}\right)\frac{4G}{r}\,, &
	 	h^+= \frac{\epsilon}{c^4}\frac{4G}{L} \ln\frac{ct-z}{r-z}\, &
	 	h^-=  \frac{\epsilon}{c^4}\frac{4G}{L} \ln\frac{ct+z}{r+z}\\
	 	\textrm{For }L \le ct-r: & h^s= \frac{1}{2c^4} \left(mc^2-2\epsilon\right)\frac{4G}{r}\, & 
	 	h^+= \frac{\epsilon}{c^4}\frac{4G}{L} \ln\frac{ct-z}{ct-L-z}\,&
	 	h^-= \frac{\epsilon}{c^4}\frac{4G}{L} \ln\frac{ct+z}{ct-L+z}
	 \end{array}
\end{equation}
with $r=(x^2+y^2+z^2)^{\frot}$. 
We find that $h_{\mu\nu}$ coincides with the Schwarzschild metric, linearized in Schwarzschild coordinates, until the start of the emission at $t=0$. 
\begin{figure}[h]
\includegraphics[width=6cm,angle=0]{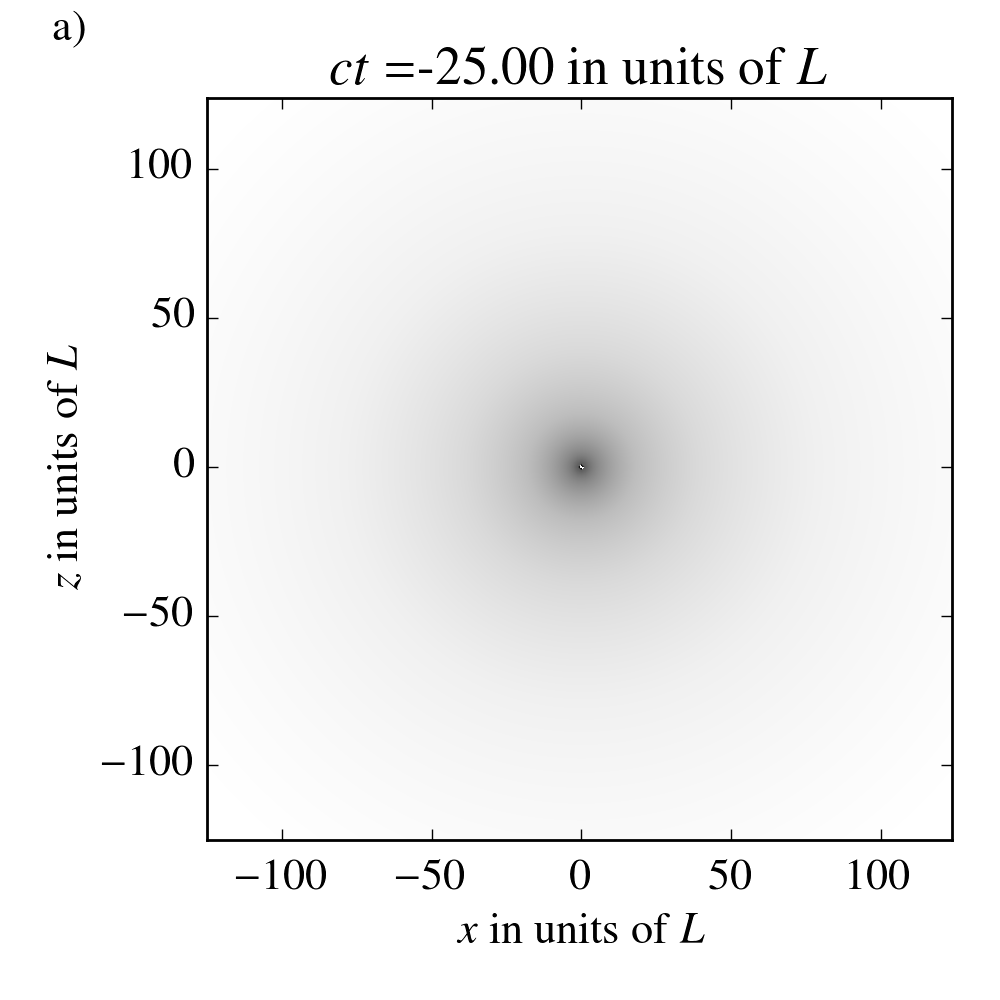}
\includegraphics[width=7.2cm,angle=0]{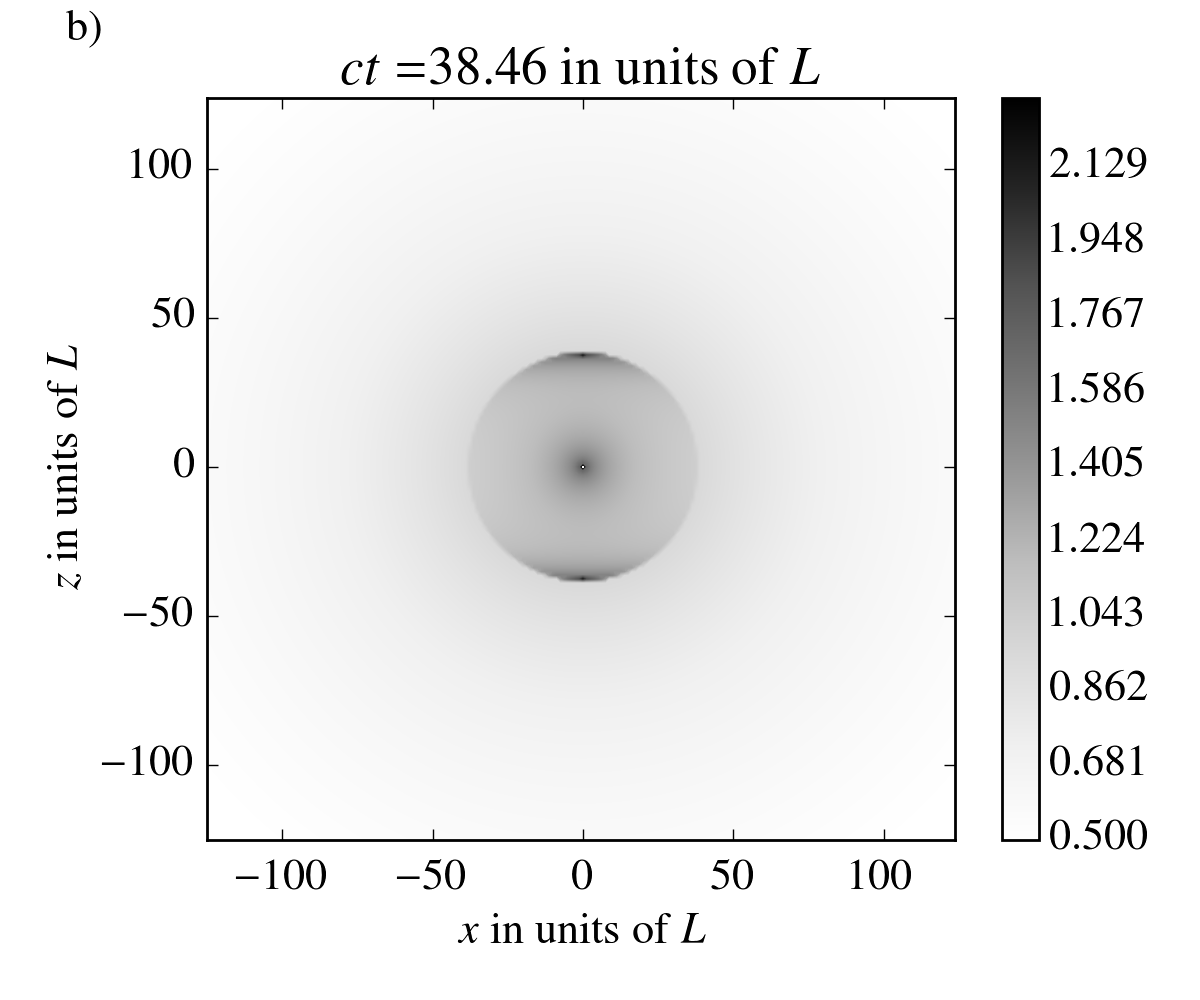}
\caption{\label{fig:hval} The plots show one component of the metric perturbation $h_{00}$ for a point particle of mass $mc^2=3\epsilon$ that emits two laser pulses of energy $\epsilon$ and length $L$ in the coordinates $(ct,x,y,z)$ in the $(x,y)$-plane for different times $t$. $h_{00}$ is normalized to units of $\kappa$ and then the logarithm of the logarithm is taken. The $mc^2$ of the emitter is chosen close to the energy of the photons $\epsilon$ such that the change in the gravitational field of the emitter due to the emission becomes visible. The model allows arbitrary masses as long as they are small enough to allow for the use of linearized gravity. In a), we see the metric spherical symmetrical perturbation due to the massive point particle at rest at $(z,x)=(0,0)$. In b), we see the effect of the two light pulses on the metric expanding from the point of their emission at $(z,x)=(0,0)$ after the emission.
%In c) and d), the field has further expanded.
}
\end{figure}
By direct calculation, it can be checked easily that (\ref{eq:gmunudecay}) fulfills the Lorentz gauge condition (\ref{eq:lorentzgauge}).
%We can check that (\ref{eq:gmunudecay}) fulfills the Lorentz gauge condition 
%(\ref{eq:lorentzgauge}). From (\ref{eq:lorentzgauge}), we obtain the three conditions
%\begin{eqnarray}\label{eq:conddecay}
%	\nonumber 2\partial_0  h_s + (\partial_0+\partial_z)h_+ + (\partial_0-\partial_z) h_- &=& 0\\
%	-(\partial_0+\partial_z)  h_+ + (\partial_0-\partial_z) h_- &=&0\,.
%\end{eqnarray}
%Direct calculation of the derivatives gives us for $0 < ct-r < L$
%\begin{eqnarray}
%	\nonumber 2\partial_0  h_s&=&-\frac{8\epsilon G}{c^4}\frac{1}{r \,L}\\
%	 (\partial_0+\partial_z) h_+ &=& \frac{4\epsilon G}{c^4}\frac{1}{r \,L}=(\partial_0-\partial_z) h_-\,
%\end{eqnarray}
%and zero everywhere else. 
In the next section, we will investigate the physical effect due to the emission we modeled.

\section{Curvature and tidal forces}
\label{sec:curvature}

%\begin{figure}[h]
%\hspace{1cm}
%\includegraphics[width=10cm,angle=0]{geodesic_dev.pdf}
%\caption{The distance between two geodesics changes due to curvature. 
% \label{fig:geodesic_dev}}
%\end{figure}
It is convenient to investigate the Riemann curvature tensor to find out if and where a physical effect of the metric perturbation (\ref{eq:gmunudecay}), in principle, could be measured. If the Riemann curvature tensor vanishes in a spacetime region, a coordinate transformation can be found in which the metric $g$ looks like the Minkowski metric $\mathrm{diag}(-1,1,1,1)$ (see Section 13.9 in \cite{Straumann2004} and proposition 2.11 of \cite{Atiyah1973}). This means that there is no measurable physical effect.

In first order in the metric perturbation $h_{\mu\nu}$, the Riemann curvature tensor takes the form
\begin{equation}\label{eq:defriemann}
	R_{\nu\rho\sigma\alpha}=\frac{1}{2}\left(\partial_{\rho}\partial_{\sigma}h_{\nu\alpha}-\partial_{\nu}\partial_{\sigma}h_{\rho\alpha}-\partial_{\rho}\partial_{\alpha}h_{\nu\sigma}+\partial_{\alpha}\partial_{\nu}h_{\rho\sigma}\right)\,.
\end{equation}
$R_{\nu\rho\sigma\alpha}$ has only $20$ independent components. This can be seen from its symmetries $R_{\nu\rho\sigma\alpha}=-R_{\rho\nu\sigma\alpha}=-R_{\nu\rho\alpha\sigma}$ and $R_{\nu\rho\sigma\alpha}=R_{\sigma\alpha\nu\rho}$ and the Bianchi identity $R_{\nu\rho\sigma\alpha}+R_{\nu\sigma\alpha\rho}+R_{\nu\alpha\rho\sigma}=0$ it fulfills. Due to these symmetries, the Riemann tensor is invariant under linearized coordinate transformations $x^\mu\rightarrow x^\mu+\xi^\mu$, where $\xi^\mu$ is assumed to be small, and terms of higher than linear order in $\xi^\mu$ are neglected. 

The Riemann tensor ${R^{\mu}}_{\rho\sigma\alpha}$ has a direct physical and geometrical interpretation. It appears naturally in the geodesic deviation equation for the relative acceleration between two infinitesimally close geodesics $\gamma(\lambda)$ and $\gamma'(\lambda)=\gamma(\lambda)+s(\lambda)$ parameterized by $\lambda$: %(see figure \ref{fig:geodesic_dev}):
\begin{equation}\label{eq:geodesicdev}
	a^\mu=\frac{D^2s^\mu}{d\lambda^2}={R^{\mu}}_{\rho\sigma\alpha}(x)\dot{\gamma}^\rho \dot{\gamma}^\sigma s^\alpha\,,
\end{equation}
where $s$ is the separation vector between the geodesics and $D/d\lambda=\dot\gamma^\mu \nabla_\mu$ is the covariant derivative along the geodesic $\gamma(\lambda)$. Equation (\ref{eq:geodesicdev}) can be interpreted as the effect of tidal forces on neighboring test particles.

The linearized Riemann curvature tensor (\ref{eq:defriemann}) corresponding to a linear combination of metric perturbations can be written as a linear combination of curvature tensors. Therefore, we obtain
\begin{equation}\label{eq:curvaturesep}
	R_{\mu\nu\rho\sigma}=	R^s_{\mu\nu\rho\sigma} + R^+_{\mu\nu\rho\sigma} + R^-_{\mu\nu\rho\sigma}
\end{equation}
with the curvature tensors corresponding to $h^s_{\mu\nu}$, $h^+_{\mu\nu}$ and $h^-_{\mu\nu}$, respectively. Expressions for the non-vanishing, independent components of the three terms in equation (\ref{eq:curvaturesep}) are given in Appendix A. Using the Lorentz gauge condition (\ref{eq:lorentzgauge}), we can write the only non-vanishing, independent components of the whole curvature as
\begin{equation}
	\begin{array}{lllllll}
		 R_{aiaj}&=&-\frac{1}{2}\partial_i\partial_j(h^s + h^- + h^+)-\frac{1}{2}\delta_{ij}\partial_a^2 h^s&\quad & R_{xyxy}&=&-\frac{1}{2}\,\left(\partial_x^2+\partial_y^2 \right)h^s\\
		 R_{0izj}&=&-\frac{1}{2}\partial_i\partial_j(h^- - h^+)-\frac{1}{2}\,\delta_{ij}\partial_z\partial_0 h^s &\quad & R_{0zai}&=& -\frac{1}{2} \partial_i\partial_a h^s \\
		  & & & & R_{0z0z}&=&\frac{1}{2}\,(\partial_0^2-\partial_z^2) h^s 
	\end{array}
\end{equation}
%\begin{eqnarray}
%	\nonumber R_{0z0z}&=&\frac{1}{2}\,(\partial_0^2-\partial_z^2) h^s\\
%	\nonumber R_{0z0i}&=& -\frac{1}{2} \partial_i\partial_z h^s\\
%	R_{0zzi}&=& -\frac{1}{2} \partial_i\partial_0 h^s\\
%	\nonumber R_{0i0j}&=&-\frac{1}{2}\partial_i\partial_j(h^s + h^- + h^+)-\frac{1}{2}\delta_{ij}\partial_0^2 h^s\\
%	\nonumber R_{0izj}&=&-\frac{1}{2}\partial_i\partial_j(h^- - h^+)-\frac{1}{2}\,\delta_{ij}\partial_z\partial_0 h^s\\
%	\nonumber R_{zizj}&=&-\frac{1}{2}\partial_i\partial_j(h^s + h^- + h^+)-\frac{1}{2}\delta_{ij}\partial_z^2 h^s\\
%	R_{xyxy}&=&-\frac{1}{2}\,\left(\partial_x^2+\partial_y^2 \right)h^s\,,
%\end{eqnarray}
where $a=0,z$ and $i,j\in \{x,y\}$.
We see that $R_{0z0z}$, $R_{0z0i}$ and $R_{0zzi}$ are of the order of derivatives of $h^s$. In order to evaluate them we must specify the creation process. This means, for example, that we cannot predict the deflection of a test particle due to one of the laser pulses along the $z$-axis in general, without first specifying the context of the laser pulse emission. The necessity to specify the context can be seen by investigating the geodesic deviation equation (\ref{eq:geodesicdev}) \cite{Raetzel2016pulse}. For expressions for the geodesic deviation due to the emitting particle and the two laser pulses see Appendix B.

\begin{figure}[h]
\includegraphics[width=6.5cm,angle=0]{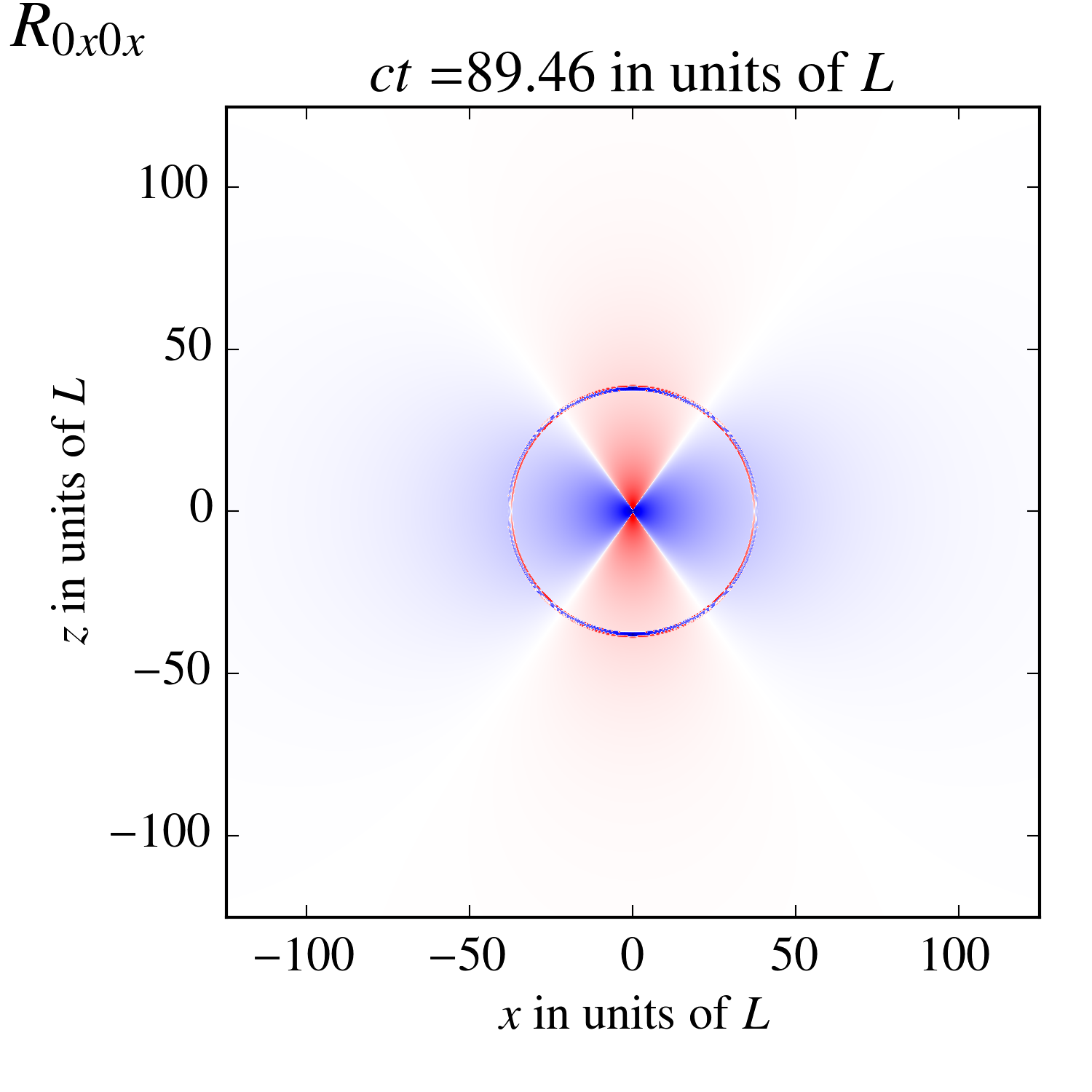}
\includegraphics[width=6.5cm,angle=0]{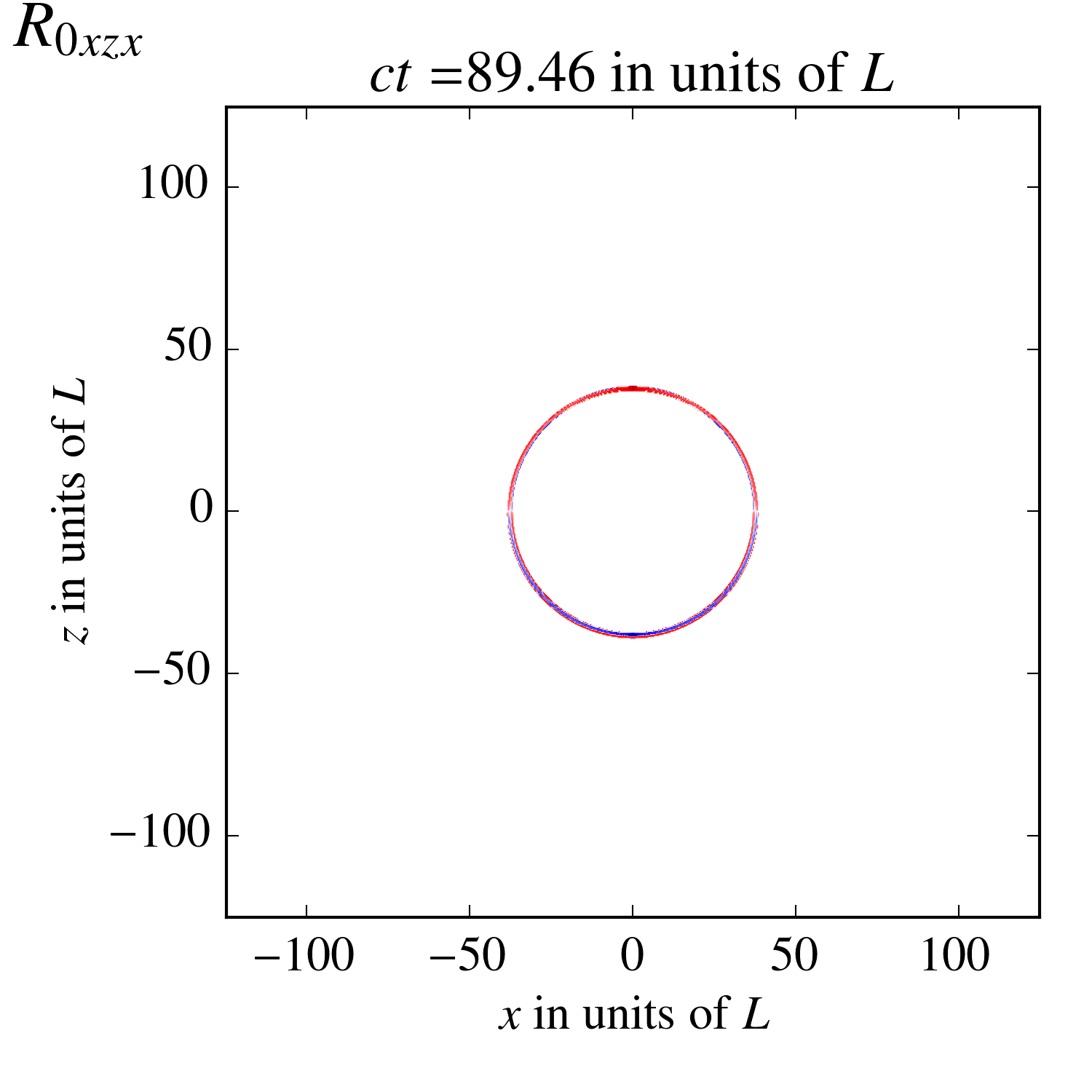}
\includegraphics[width=6.5cm,angle=0]{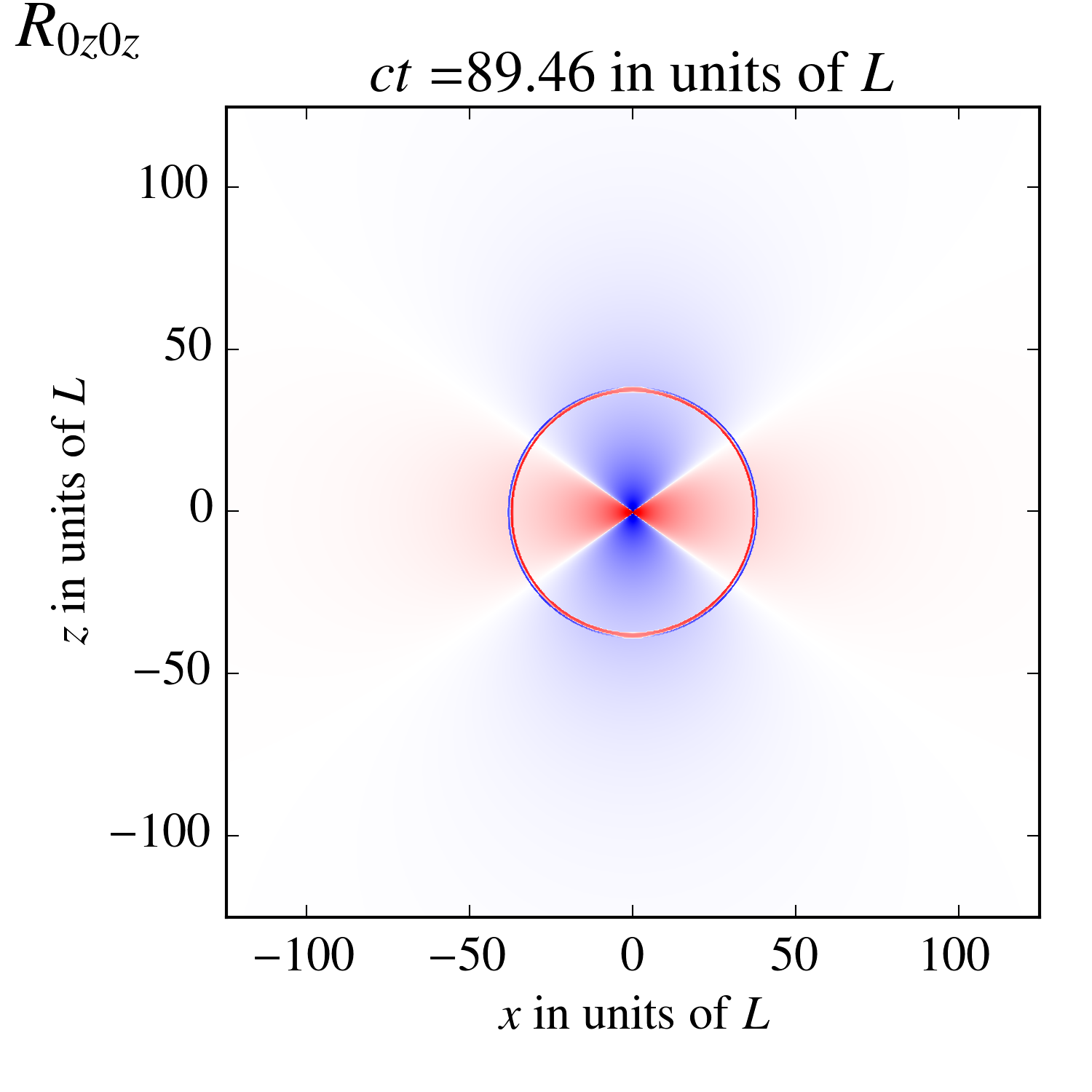}
\includegraphics[width=6.5cm,angle=0]{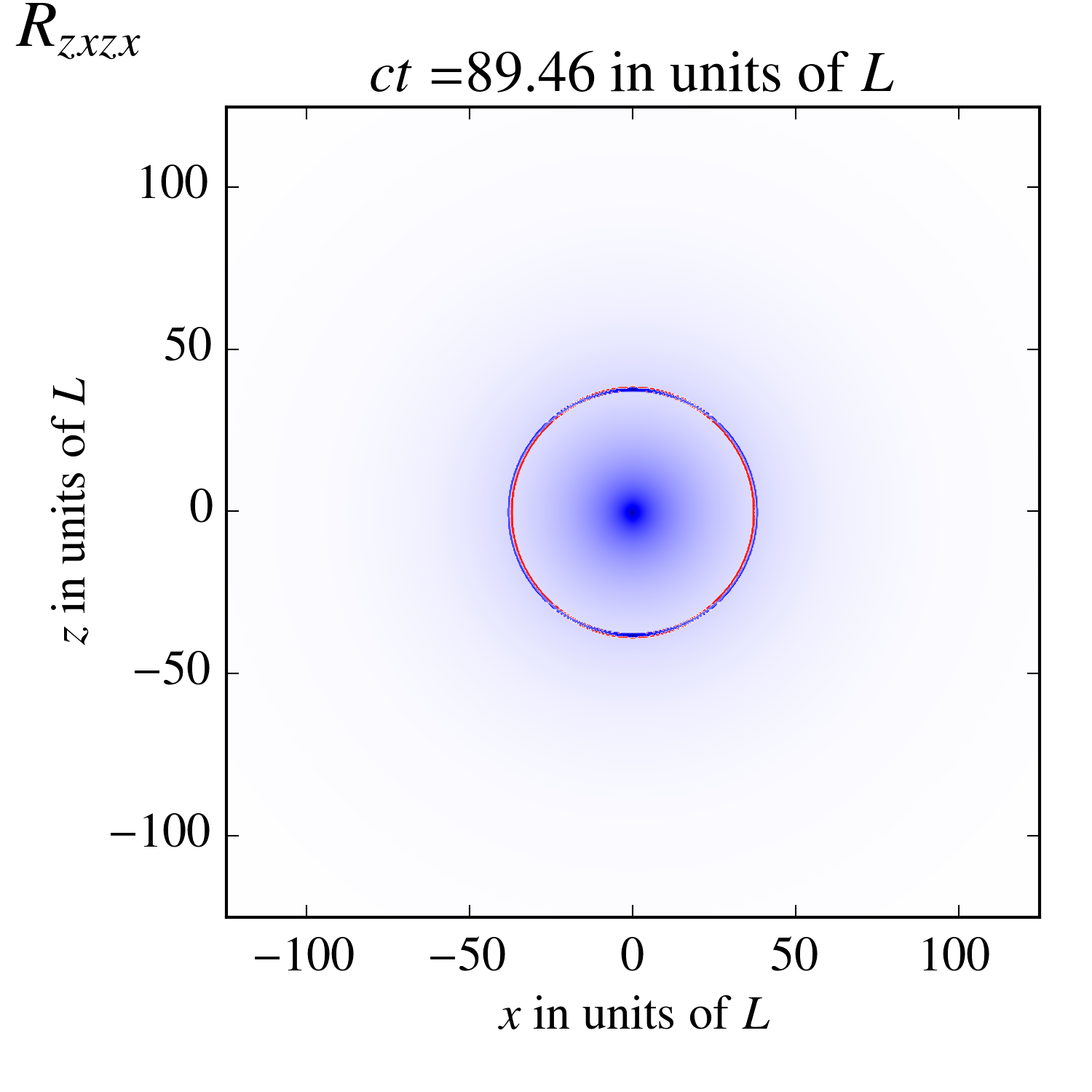}
\caption{\label{fig:curve} The plots show the curvature components $R_{0x0x}$, $R_{0xzx}$, $R_{0z0z}$ and $R_{zxzx}$ for the metric perturbation $h_{\mu\nu}$ induced by a point particle of mass $mc^2=3\epsilon$ that emits two laser pulses of energy $\epsilon$ in the coordinates $(ct,x,y,z)$ in the $(x,z)$-plane for different times $t$. The logarithm of value of the curvature components is encoded in the opacity of the color. Red is a negative value of curvature component and blue a positive value. White stands for zero.}
\end{figure} 
Plots for some of the curvature components are given in Figure \ref{fig:curve}. The components presented in the plots as well as all other components share the following features: Inside a sphere, which is expanding with the speed of light, the curvature is that of a point particle with the reduced mass $m-2\epsilon/c^2$. This sphere corresponds to the end of the emission process, when the point particle hast lost the mass $2\epsilon/c^2$, due to the emission of the laser pulses. The sphere is the intersection of the spatial plane of constant $ct$ with the post-emission cone (see Figure \ref{fig:Regions}).
We can distinguish two further regions. The first is the intersection of the $ct=const.$-plane with the pre-emission cone. Here, the gravitational field is that of the emitter with mass $m$. The spherical shell between the first and the second spherical regions is the intersection of the $ct=const.$-plane with the emission cone. It is only here that we find a gravitational effect of the laser pulses since only here they contribute to the curvature. We see that the gravitational field of light (corresponding to our model and the framework of linearized gravity) is only due to its emission. In the next section, we will investigate the acceleration of a test particle due to the metric perturbation (\ref{eq:gmunudecay}) in the emission cone.

\section{Acceleration of a test particle at rest}
\label{sec:acc}

In this section, we investigate the acceleration experienced by a test particle at rest with respect to the emitter. In the coordinates $(ct,x,y,z)$, the geodesic equation governing the trajectories $\gamma$ of freely falling test particles is given in first order in the metric perturbation as
\begin{equation}\label{eq:geodesic}
 \ddot{\gamma}^\mu=-\left(\eta^{\mu\nu}\left(\partial_\rho h_{\nu\sigma} -\frac{1}{2}\partial_\nu h_{\rho\sigma}\right)-h^{\mu\nu}\left(\partial_\rho \eta_{\nu\sigma} -\frac{1}{2}\partial_\nu \eta_{\rho\sigma}\right)\right)\dot{\gamma}^\rho \dot{\gamma}^\sigma\,.
\end{equation}	
%For the metric perturbation (\ref{eq:gmunudecay}), we find
%\begin{eqnarray}
%	\ddot x^0&=&-\dot x^0\left( \dot x^\mu\partial_\mu -\frac{1}{2}\dot x^0\partial_0\right)h_{00} - \dot x^3\left( \dot x^\mu\partial_\mu -\dot x^0\partial_0\right)h_{03} + \frac{1}{2}(\dot x^3)^2\partial_0 h_{33}\\
%	\ddot x^i&=&-\frac{1}{2}\left((\dot x^0)^2\partial_i h_{00}+2\dot x^0 \dot x^3\partial_i h_{03}+(\dot x^3)^2\partial_i h_{33}\right)\\
%	\ddot x^3&=& -\frac{1}{2}(\dot x^0)^2\partial_3 h_{00} +\dot x^0\left( \dot x^\mu\partial_\mu -\dot x^3\partial_3\right)h_{03} + \dot x^3\left( \dot x^\mu\partial_\mu -\frac{1}{2}\dot x^3\partial_3\right)h_{33}
%\end{eqnarray}
We will only consider terms in the metric perturbation that contribute to the curvature. All other terms can be canceled by a linearized coordinate transformation which changes the coordinates only by terms that are of the order $\kappa=\frac{4 G\epsilon}{c^4L}$. Hence, the effect of this coordinate transformation on equation (\ref{eq:geodesic}) is of order $\kappa^2$ and can be neglected as a contribution of higher order. Then, in the emission cone, the functions $h^s$, $h^+$ and $h^-$ are given as
\begin{equation}\label{eq:hfunccurv}
	h^s=\kappa \frac{\mu - ct + r}{r}\,,\quad h^+= -\kappa(\ln r +\ln(1-cos\theta))\,, \quad h^-= -\kappa(\ln r + \ln(1+cos\theta))\,,
\end{equation}
expressed in terms of the spherical coordinates $z=r\cos\theta$, $x=r\cos\phi\sin\theta$ and $y=r\sin\phi\sin\theta$ and using the definition $\mu=\frac{Lmc^2}{2\epsilon}$. In these coordinates, the Minkowski metric has the form $\eta=\mathrm{diag}(-1,1,r^2,r^2\sin^2\theta)$.

For a test particle at rest we have $\dot\gamma=(c,0,0,0)$, and the geodesic equation (\ref{eq:geodesic}) becomes
\begin{equation}\label{eq:geodesicrest}
 \ddot{\gamma}^\mu=-c^2\eta^{\mu\nu}\left(\partial_0 h_{\nu0} -\frac{1}{2}\partial_\nu h_{00}\right)\,.
\end{equation}	
This means that only $\partial_a h_{00}$ and $\partial_0 h_{a0}$ with $a=x,y,z$ contribute to the spatial part of $\ddot{\gamma}$. From the equations (\ref{eq:hfunccurv}), we see that the metric perturbation is time independent, and the contributions of $\partial_0 h_{a0}$ vanish.

It is interesting to consider the contributions of the massive particle and the laser pulses separately. We will start with the contribution of the massive point particle. We find from equation (\ref{eq:geodesicrest}) that a test particle at rest experiences a radial acceleration
\begin{equation}\label{eq:ddotr}
	\ddot r^s = \frac{c^2}{2}\partial_r h^s = -\frac{c^2}{2}\kappa \frac{\mu-ct}{r^2}
\end{equation}
inside the emission cone due to the massive point particle. Interestingly, it contains an acceleration away from the origin as we will see in the following. In the emission cone, $ct$ lies between $r$ and $r+L$. We can define the retarded time $t_\mathrm{ret}=t - r/c$ with $0\le ct_\mathrm{ret}\le L/c$ in the emission cone, and we find
\begin{equation}\label{eq:pointpartacc}
	\ddot r^s = -\frac{c^2}{2}\kappa \left( \frac{\mu-ct_\mathrm{ret}}{r^2} - \frac{1}{r} \right) = - \frac{G}{r^2}\left(m - \frac{2\epsilon}{c^2} \frac{ct_\mathrm{ret}}{ L}\right) + \frac{2G\epsilon}{c^2L}\frac{1}{r}
\end{equation}
The first term in (\ref{eq:pointpartacc}) is an acceleration proportional to the mass of the point particle $m - \frac{2\epsilon}{c^2}\frac{ct_\mathrm{ret}}{L}$ at the retarded time $t_\mathrm{ret}$. This acceleration is proportional to $1/r^2$ and corresponds to the Newtonian attraction of the massive point particle. It gradually decreases as the emission cone passes the position of the test particle in the retarded time $t_\mathrm{ret}=ct-r$ corresponding to the position of the test particle. 

The second part gives rise to an acceleration away from the origin. This acceleration is proportional to 1/r, where r is the distance from the point particle. We can call it a gravitational induction force as it is associated with the change of the gravitational field due to the mass loss of the emitter.

For the contribution of the two laser pulses, we find from equation (\ref{eq:geodesicrest}) for the spatial part of the acceleration
\begin{equation}\label{eq:laseracc}
	\left(\begin{matrix} \ddot r^p \\ r\ddot \theta^p \end{matrix}\right) =\frac{c^2}{2} \left(\begin{matrix} \partial_r (h^+ + h^-) \\ \frac{1}{r}\partial_\theta (h^+ + h^-) \end{matrix}\right) =  -\frac{4G\epsilon}{c^2 L } \frac{1}{\rho}\left(\begin{matrix} \sin\theta \\ \cos\theta \end{matrix}\right)\,,
\end{equation}
where $\rho=\sqrt{x^2+y^2}=r\sin\theta$. This acceleration always points towards the $z$-axis. 
For long lifetimes after the pulse emission or close to its trajectory, the acceleration in equation (\ref{eq:laseracc}) coincides with the attraction experienced by a massive test particle at rest due to a single laser pulse.

Note that the absolute value of the radial part of the acceleration due to the laser pulses in equation (\ref{eq:laseracc}) is exactly twice as large as the induction force (\ref{eq:pointpartacc}) due to the mass change of the point particle, the second term in equation (\ref{eq:pointpartacc}). Since they are of opposite sign, the half of the radial acceleration due to the laser pulses is canceled and the total radial acceleration is
\begin{equation}\label{eq:radacc}
	\ddot r =\ddot r^s + \ddot r^p =- \frac{G}{r^2}\left(m - \frac{2\epsilon}{c^2} \frac{ct_\mathrm{ret}}{ L}\right)- \frac{2G\epsilon}{c^2L}\frac{1}{r}\,.
\end{equation}
Hence, the non-Newtonian part of the radial acceleration, proportional to $\frac{1}{r}$, is attractive. Since the emission is not isotropic, there is an acceleration in the $\theta$-direction. It is only due to the pulses and given as
\begin{equation}\label{eq:rddottheta}
	r\ddot \theta = r\ddot \theta^p = \frac{c^2}{2}\frac{1}{r}\partial_\theta h_{00}  =-\frac{c^2\kappa}{2r} \left(\frac{\sin{\theta}}{1-\cos\theta}-\frac{\sin{\theta}}{1+\cos\theta}\right) = -\frac{4G\epsilon}{c^2 L }\frac{1}{r} \cot\theta\,.
\end{equation} 
In the next section, we want to investigate the acceleration of a massless test particle witnessing the emission process.

\section{Acceleration of massless test particles}
\label{sec:masslesstest}

For a massless test particle traveling in the -$z$-direction, the situation is very different than for a particle at rest. The $4$-velocity vector is $\dot\gamma=(c,0,0,-c)$. Hence, we find with the geodesic equation (\ref{eq:geodesic})
\begin{equation}\label{eq:vecxhsh+}
	\ddot{\vec{\gamma}}=c^2\left[\left(\begin{matrix} \partial_\rho h^s \\ \partial_0 h^s \end{matrix}\right) +2 \left(\begin{matrix}
	\partial_\rho h^+ \\ \partial_0 h^+
\end{matrix}	\right)  \right]\,,
\end{equation}
where we defined the acceleration vector as $\ddot{\vec{\gamma}}:=(\ddot \rho,\ddot z)$ in the cylindrical coordinates $(ct,\rho,\phi,z)$ with $x=\rho\cos\phi$ and $y=\rho\sin\phi$. $\ddot \rho$ expresses the transversal acceleration and $\ddot z$ the longitudinal acceleration with respect to the $z$-axis. The second laser pulse does not contribute to the acceleration since the massless test particle propagates parallel to the second laser pulse and into the same direction.

Again neglecting all terms that do not contribute to the curvature, we have
\begin{eqnarray}
	h^s &=& \kappa \frac{\mu - ct + r}{r}\\
	h^+ &=& -\kappa \ln(r-z)\,
\end{eqnarray}
in the emission cone. Then, evaluation of the expressions in (\ref{eq:vecxhsh+}) gives 
\begin{eqnarray}
	 \ddot{\vec{\gamma}}&=&-\frac{\kappa c^2}{r}\left(\begin{matrix} \left(\frac{\mu - ct}{r^2} + 2\frac{1}{r-z}\right)\rho \\ 1\end{matrix}\right)
\end{eqnarray} 
Note, that the contribution of the laser pulse to the acceleration is only transversal. With the parameterization $ct_\mathrm{ret}=ct-r$ in the emission cone we find
\begin{eqnarray}\label{eq:restacc}
	 \ddot{\vec{\gamma}}&=&-\frac{2G}{r^2}\left(m - \frac{2\epsilon}{c^2} \frac{ct_\mathrm{ret}}{L}\right)\left(\begin{matrix} \frac{\rho}{r} \\ 0 \end{matrix}\right)-\frac{4G\epsilon}{c^2L }\frac{1}{r}\left(\begin{matrix}\frac{r+z}{r-z}\frac{\rho}{r} \\ 1 \end{matrix}\right)\,.
\end{eqnarray} 
In equation (\ref{eq:restacc}), there are two terms with differing dependence on the radial distance $r$. The term that is proportional to $1/r$ is due to the emission of radiation and the change of mass of the emitter.  The first term in (\ref{eq:restacc}) is the gravitational acceleration of light by a massive object of mass $\left(m - \frac{2\epsilon}{c^2} \frac{ct_\mathrm{ret}}{L}\right)$. The factor $2$ in comparison to the Newtonian force experienced by a massive particle at rest given in equation (\ref{eq:radacc}) is
known from general relativistic effects like gravitational lensing.

For very small $\rho/r$ and positive $z$, we find that only the gravitational effect of the laser pulse contributes, and we obtain
\begin{eqnarray}\label{eq:restaccclose}
	 \ddot{\vec{\gamma}}
&=&-\frac{16G\epsilon}{c^2L}\frac{1}{\rho} \left(\begin{matrix}
1 \\ 0
\end{matrix}	\right) 
\end{eqnarray} 
%This coincides with the result presented in \cite{Raetzel2016pulse} for a single laser pulse far from its emission and close to its trajectory.

\section{Isotropic emission and the relation to the Vaidya metric}
\label{sec:compvaidya}

In the case of the emission of two counter propagating laser pulses, light is emitted with directional preference. In this section, we will discuss the isotropical case and its relation to the metric due to a spherically symmetrical, isotropically radiating massive object; the Vaidya metric.

In the limit of infinitely many laser pulses of infinitely small energy that are emitted isotropically in all spatial directions, the energy momentum tensor becomes
\begin{equation}\label{eq:energymomentumlinvaidya}
	T_{\mu\nu}(x)= \left\{\begin{array}{ccc}
		 0 &:& ct-r\le ct\\
		 \frac{P}{ 4\pi r^2c}l_\mu(x) l_\nu(x)  &:& 0 < ct-r < L\\ 
		 0 &:& L \le ct-r
	\end{array}\right.
\end{equation} 
where $P$ is the total radiation power and $l_\mu(x)$ is the four momentum of a light ray that is moving radially outwards at the spacetime point $x$ and that fulfills the normalization condition $l_0=-1$. We define the retarded time $u=ct-r$ and the advanced time $v=ct+r$. In the set of coordinates $(u,v,\theta,\phi)$, the only non-vanishing component of the energy momentum tensor (\ref{eq:energymomentumlinvaidya}) is $T_{uu}=\frac{P}{4\pi r^2c}$. We define the mass of the emitter as $M(u)=m-Pu/c^3$. Then, we can identify $P$ with the rate of mass loss of the central particle $-c^3\partial_u M(u)$, and we find
\begin{equation}\label{eq:emvaidya}
	T_{uu}=-\frac{c^2\partial_u M(u)}{4\pi r^2}\,.
\end{equation}
A solution of the full Einstein equations to this energy momentum tensor for general monotonously decreasing functions $M(u)$ is the retarded Vaidya metric. It was derived as the spacetime induced by a radiating spherically symmetric massive object in\cite{Vaidya1951} by Vaidya. It is given by the line element (see \cite{Lindquist1965})
\begin{equation}\label{eq:vaidya}
	ds^2=-\left(1-\frac{2GM(u)}{c^2r}\right)du^2-2dudr+r^2d\Omega^2=\frac{2GM(u)}{c^2r}du^2+ds^2_{\mathbb{M}}\,,
\end{equation}
where $d\Omega^2=d\theta^2+\sin^2\theta d\varphi^2$ and $ds^2_{\mathbb{M}}$ is the line element of Minkowski space. Note that the Vaidya metric becomes the Schwarzschild metric for constant mass, which takes its standard form $ds^2=-(1-\frac{2GM}{c^2r})dc\bar{t}^2+(1-\frac{2GM}{c^2r})^{-1}dr^2+r^2d\Omega^2$ using the time coordinate $c\bar t=u+r+2M\ln(r/2M-1)$.
\begin{figure}[h]
%\hspace{2cm}
\includegraphics[width=10cm,angle=0]{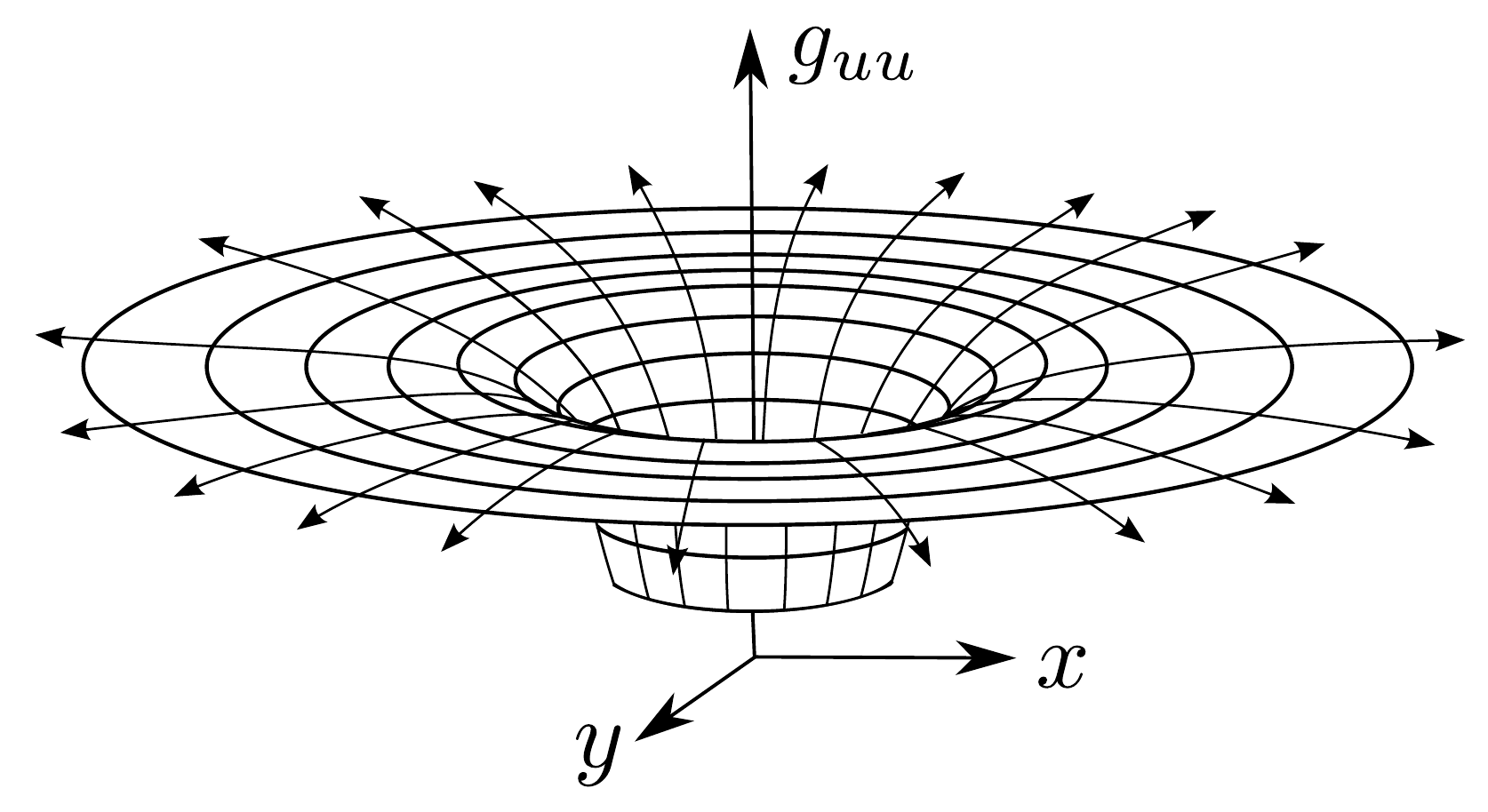}
\caption{The Vaidya metric can be interpreted as a Schwarzschild metric with changing mass. The dependence of the mass term on the retarded time leads to an everywhere non-zero energy momentum tensor which can be interpreted as a radially outward directed flux of electromagnetic radiation moving in the Schwarzschild background (represented by radial arrows in the picture).
 \label{fig:vaidya}}
\end{figure}
%Also, for constant mass, the Schwarzschild time can be defined as
%\begin{equation}\label{eq:retardedtime}
%ct=u+r+\frac{2GM}{c^2}\ln\left(\frac{c^2r}{2GM}-1\right)\,
%\end{equation}
%which for small masses reduces to $ct=u+r$. 
From the second expression for the line element in equation (\ref{eq:vaidya}) we see that a linearized version of the Vaidya metric is given by the metric perturbation $h^V_{\mu\nu}$ with the only non-vanishing component
\begin{equation}\label{eq:vaidyalin}
	h^V_{uu}=\frac{2GM(u)}{c^2r}
\end{equation}
where $\chi(u)$ is the characteristic function encoding the envelope of the pulse. The metric perturbation $h^V_{\mu\nu}$ presented in equation (\ref{eq:vaidyalin}) does not fulfill the Lorentz gauge condition. Therefore, we are not able to derive it using the method of retarded potentials which we employed in Section \ref{sec:pulse}. However, the discrepancy is only due to a different choice of coordinates. Thus, the linearized Vaidya metric (\ref{eq:vaidyalin}) can really be seen as the isotropic version of the pulse emission metric $\eta_{\mu\nu}+h_{\mu\nu}$ with $h_{\mu\nu}$ given in equation (\ref{eq:gmunudecay}).

The equations of motion for a test particle in the full Vaidya metric (\ref{eq:vaidya}) can be derived from the metric $g_{\mu\nu}$ encoded in the line element $ds^2=g_{\mu\nu}dx^\mu dx^\nu$ using the geodesic equation
\begin{equation}\label{eq:geodesiceq}
\ddot{\gamma}^\mu = -{\Gamma^\mu}_{\rho\sigma}\dot{\gamma}^\rho\dot{\gamma}^\sigma
\end{equation}
with subsidiary conditions $g_{\mu\nu}\dot{\gamma}^\mu\dot{\gamma}^\nu=-1$ for massive test particles, and $g_{\mu\nu}\dot{\gamma}^\mu\dot{\gamma}^\nu = 0$ for massless test particles. In equation (\ref{eq:geodesiceq}), the dot indicates the derivative with respect to the curve parameter, which is proper time $\tau$ for the time-like geodesics of massive particles. For the null-geodesics of massless particles, we will use coordinate time $t$. ${\Gamma^\mu}_{\rho\sigma}$ are the Christoffel symbols, ${\Gamma^{\mu}}_{\rho\sigma} = \frac{1}{2} g^{\mu\nu}\left(\partial_\rho g_{\sigma\nu} + \partial_{\sigma} g_{\nu\rho} -\partial_\nu g_{\rho\sigma}\right)$.

From the line element (\ref{eq:vaidya}), we obtain the equations of motion in spherical coordinates
\begin{eqnarray}\label{eq:geodesicdevexpl}
 \nonumber  \ddot{\gamma^r} & = & -\frac{GM(u)}{c^2r^2}\left(\left(1-\frac{2GM(u)}{c^2r}\right)-r\partial_u\ln M(u)\right)\left(\dot\gamma^u\right)^2-\frac{2GM(u)}{c^2r^2}\dot\gamma^u\dot\gamma^r +\\
\nonumber  && + \,r\left(1-\frac{2GM(u)}{c^2r}\right)\left(\left(\dot\gamma^\theta\right)^2+\sin^2\theta\left(\dot\gamma^\phi\right)^2\right)
   \\
   \ddot{\gamma^\theta} & = & -\frac{2}{r}\dot\gamma^\theta\dot\gamma^r + \sin\theta\cos\theta \left(\dot\gamma^\phi\right)^2 
   \\
 \nonumber  \ddot{\gamma^\phi} & = & -\left(\frac{2}{r}\dot\gamma^r + \cot\theta\dot\gamma^\theta \right)\dot\gamma^\phi
\end{eqnarray}
and the subsidiary conditions
\begin{equation}\label{eq:subsed}
   -\left(1-\frac{2GM(u)}{c^2r}\right) \left(\dot\gamma^u\right)^2 - 2\dot\gamma^u\dot\gamma^r + r^2 \left(\left(\dot\gamma^\theta\right)^2+\sin^2\theta\left(\dot\gamma^\phi\right)^2\right) =
   \left\{\begin{array}{ccl}
   -1 & \quad & \mbox{for time-like geodesics}
   \\
   0 & \quad & \mbox{for null-geodesics}
   \end{array}\right.\,.
\end{equation}
%For radially moving massless test particles, $\dot\gamma^\theta=0=\dot\gamma^\phi$, the equations reduce to $\dot\gamma^u=0$ and $\ddot\gamma^r=0$ for outward propagating particles and $\dot\gamma^u=-2\left(1-\frac{2GM(u)}{c^2r}\right)^{-1}\dot\gamma^{r}$ and $\ddot\gamma^r=\frac{4G\partial_u M(u)}{c^2r}\left(1-\frac{2GM(u)}{c^2r}\right)^{-2}\left(\dot\gamma^r\right)^2$ for inward propagating particles. 
For a massive particle initially at rest, we obtain from the subsidiary condition $\left(\dot\gamma^u\right)^2=\left(1-\frac{2GM}{c^2r}\right)^{-1}$ and, hence, 
\begin{equation}\label{eq:restaccvaidya}
	\ddot{\gamma^r}  =  -\frac{GM(u)}{c^2r^2} + \left(1-\frac{2GM(u)}{c^2r^2}\right)^{-1}\frac{G\partial_u M(u) }{c^2r}
\end{equation} 
For $M(u)=m-Pu/c^3$ and small mass $m$, equation (\ref{eq:restaccvaidya}) becomes  equation (\ref{eq:radacc}) which we derived for the emission of two counter-propagating laser pulses. The first term in equation (\ref{eq:restaccvaidya}) is the Newtonian gravitational acceleration due to a spherically symmetric massive object whose mass decreases as a function of the retarded time $u$. The second term decreases only as $1/r$ and is proportional to the radiation power of the spherical object $P=-c^3\partial_u M$. The term "gravitational induction force" for the second term was coined in \cite{Lindquist1965}. As we explained in Section \ref{sec:acc}, it contains contributions from the massive body and from the radiation that cancel partially. The contribution due to the mass loss of the massive body leads to an acceleration away from the body. The contribution of the radial radiation leads to an inward acceleration that is twice as strong as the acceleration due to the mass loss.

Due to the different dependence of the first and second term in equation (\ref{eq:restaccvaidya}) on the distance to the radiating object $r$, the second term of equation (\ref{eq:restaccvaidya}) becomes dominant for distances larger than $r_0=M/\partial_u M$. At this distance, the induction force is $G(\partial_u M)^2/Mc^2$. Hence, there needs to be a very high emission rate for the induction force to be significant. In \cite{Lindquist1965}, it is argued that the gravitational induction plays a role for radiating stars, when at all, only in catastrophic phases of gravitational collapse. For a decaying particle, however, $r_0$ is $L=cT$, where $T$ is the duration of the emission process. If the decay process takes a femtosecond, $r_0$ is only of the order $100\mathrm{nm}$. Hence, already very close to the decaying particle, the induction force becomes the dominant force during the decay. For a rubidium atom emitting a femtosecond light pulse at $780\mathrm{nm}$, we find that $r_0=15\mathrm{km}$. Hence, the radial induction force is much weaker than the Newtonian gravitational force for all relevant distances to the atom. In the an-isotropic case of the emission of two counter-propagating laser pulses, there is a non-spherical-symmetrical part of the force, which is only due to the laser pulses. As can be seen in the equations (\ref{eq:radacc}), (\ref{eq:rddottheta}) and (\ref{eq:restaccclose}), this is the only significant force at points far from the emitter but close to the trajectory of one of the pulses.

\section{Conclusions}
\label{sec:conclusions}

In this article, the situation of two counter-propagating laser pulses emitted from a massive point particle was considered. The corresponding metric perturbation in the framework of linearized gravity and the corresponding curvature were derived. It was shown that the curvature is that of a massive point particle at all spacetime points lying in the causal future of the end of the emission process and in the causal past of the beginning of the emission process. It was concluded that the laser pulses only contribute to the curvature during their emission and their absorption. This is in agreement with the results presented in \cite{Raetzel2016pulse}, where only one pulse was considered and the gravitational effect of the emitter was neglected. In contrast to the model presented in the former article, in the model presented in this article, the emitter itself is taken into account, and the continuity equation of general relativity is fulfilled.

The acceleration of test particles at rest and massless test particles propagating parallel to the trajectories of the two laser pulses in the gravitational field of the emitter and the two light pulses was calculated. It was found that the acceleration due the emitter can be separated into two qualitatively different terms. One is proportional to the inverse square of the distance to the emitter. This term is the Newtonian attraction due to the emitter and it decreases with its mass loss. The second part of the acceleration is proportional to the inverse of the distance to the emitter. It is the gravitational induction force due the change of mass of the emitter. 

The effect of the laser pulses on the test particles is an acceleration which is proportional to the inverse of the distance to the axis on which the two pulses propagate. It always points in the direction of this axis.

In the model presented in this article, the only condition on the mass of the emitter is that it is small enough to allow for the application of the linearized Einstein equations. Hence, the emitter could be, for example, an atom or a laser device emitting in two opposite directions. In that case, the energy corresponding to the rest mass of the point particle is much larger than the energy of the two light pulses.

Another particular case included in the model presented in this article is the equivalence of the total energy of the emitted laser pulses $2\epsilon$ and the energy contained in the rest mass of the point mass $mc^2$. This is the case of a complete decay of the massive point particle. The resulting energy momentum tensor and the corresponding metric perturbation were already presented in \cite{Voronov1973} by Voronov and Kobzarev. The authors modeled the creation of massless particles by the decay of a massive scalar particle. This can be seen as a classical model for the decay of a Higgs boson into two photons. 

The case of the isotropic emission of laser pulses from a massive point particle was discussed in Section \ref{sec:compvaidya}. It was argued that this situation corresponds to a particular Vaidya metric with a pulsed mass loss. The corresponding acceleration of massive test particles was investigated. Again, two terms with different dependence on the distance to the emitter were obtained: a Newtonian gravitational force proportional to the mass of the emitter that decreases with the distance square and a force proportional to the radiation power that decreases with the distance to the emitter. The latter was called the gravitational induction force. Since the induction force decreases more slowly than the Newtonian force, it becomes larger than the latter at some distance from the emitter. However, for emitting atoms it is much smaller than the Newtonian force for all experimentally relevant distances. For a decaying particle it becomes relevant already when very close to the particle. This agrees with the observation of \cite{Lindquist1965} that the induction force of a radiating star is only significant in catastrophic situations of gravitational collapse.

As explained above, the situation is different for the an-isotropic situation of two counter-propagating laser pulses emitted from a massive point particle which was considered in this article; there is a non-radial acceleration of test particles that is due only to the laser pulses.

%
%\section*{Acknowledgements}

\section*{Appendix A}

The only non-vanishing, independent components of the three curvature terms are
\begin{eqnarray}
	\nonumber R^+_{0z0z}&=&-\frac{1}{2}\,(\partial_0+\partial_z)^2 h^+\\
	R^+_{0z0i}=-R^+_{0zzi}&=&-\frac{1}{2} \partial_i(\partial_0+\partial_z)h^+\\
	\nonumber R^+_{0i0j}=R_{+\,zizj}=-R^+_{0izj}&=&-\frac{1}{2} \partial_i \partial_j h^+\,,
\end{eqnarray}
and
\begin{eqnarray}
	\nonumber R^-_{0z0z}&=&-\frac{1}{2}\,(\partial_0-\partial_z)^2 h^-\\
	R^-_{0z0i}=R^-_{0zzi}&=&\frac{1}{2} \partial_i(\partial_0-\partial_z)h^-\\
	\nonumber R^-_{0i0j}=R^-_{zizj}=R^-_{0izj}&=&-\frac{1}{2} \partial_i \partial_j h^-\,,
\end{eqnarray}
and
\begin{equation}
 	\begin{array}{lllllll}
		 R^s_{aiaj}&=&-\frac{1}{2}\,\left(\partial_a^2\delta_{ij}+\partial_i\partial_j \right)h^s &\quad & R^s_{0z0i}&=&-\frac{1}{2}\,\partial_i\partial_z h^s \\
		R^s_{0z0z}&=&-\frac{1}{2}\,\left(\partial_0^2+\partial_z^2 \right)h^s &\quad &   R^s_{0zzi}&=&\frac{1}{2}\,\partial_i\partial_0 h^s \\
		 R^s_{xyxy}&=&-\frac{1}{2}\,\left(\partial_x^2+\partial_y^2 \right)h^s  & & R^s_{0izj}&=&-\frac{1}{2}\,\delta_{ij}\partial_z\partial_0 h^s
	\end{array}
\end{equation}
where $a=0,z$ and $i,j=x,y$.

\section*{Appendix B}

For example, consider the case of two geodesics that start at the same time and move initially with $\dot \gamma=(\dot \gamma^0,0,\dot \gamma^y,\dot \gamma^z)$ and an infinitesimal separation $s=(0,-ds,0,0)$. We find from equation (\ref{eq:geodesicdev})
\begin{eqnarray}
    a^0&=&ds((R_{0y0x}\dot \gamma^0+R_{0yyx}\dot \gamma^y)\dot \gamma^y+(R_{0zyx}+R_{0yzx})\dot \gamma^z\dot \gamma^y+(R_{0z0x}\dot \gamma^0+R_{0zzx}\dot \gamma^z)\dot \gamma^z)\\
	a^x&=&-ds((R_{x00x}\dot \gamma^0+2R_{x0yx}\dot \gamma^y + 2R_{x0zx}\dot \gamma^z)\dot \gamma^0+2R_{xyzx}\dot \gamma^y \dot \gamma^z + R_{xyyx}(\dot \gamma^y)^2 + R_{xzzx}(\dot \gamma^z)^2)\\
	a^y&=&-ds((R_{y00x}\dot \gamma^0 + R_{y0yx}\dot \gamma^y+(R_{y0zx}+R_{yz0x})\dot \gamma^z)\dot \gamma^0 + (R_{yzyx}\dot \gamma^y + R_{yzzx}\dot \gamma^z)\dot \gamma^z)\\
	a^z&=&-ds((R_{z00x}\dot \gamma^0+(R_{z0yx}+R_{zy0x})\dot \gamma^y+R_{z0zx}\dot \gamma^z)\dot \gamma^0+ ( R_{zyyx} \dot \gamma^y + R_{zyzx}\dot \gamma^z) \dot \gamma^y)\,.
\end{eqnarray}
Only $\frac{D^2s^i}{d\tau^2}$ with $i=x,y$ is independent of the curvature terms $R_{z00j}$ and $R_{z0zj}$ for which the contributions of the metric perturbation due to the point mass cannot be neglected in general. 

\section*{References}

\bibliographystyle{ieeetr} %plainnat
\bibliography{gravphoton}

\end{document}